\documentclass{iaus}
\usepackage{graphicx}

\title[Star formation: metallicity and X-rays] 
{The impact of metallicity and X-rays on star formation}

\author[Spaans, Aykutalp, Hocuk]   
{Marco Spaans$^1$, Aycin Aykutalp$^1$, Seyit Hocuk$^1$}

\affiliation{$^1$Kapteyn Astronomical Institute, University of Groningen, \\ P.O.\ Box 800, 9700 AV Groningen, the Netherlands \\ email: {\tt spaans@astro.rug.nl} }

\pubyear{2010}
\volume{270}  
\pagerange{1--4}
\jname{Computational Star Formation}
\editors{J. Alves, B. Elmegreen, J. P. Girart \& V. Trimble, eds.}
\begin{document}

\maketitle

\begin{abstract}

Star formation is regulated through a variety of feedback processes. In 
this study, we treat feedback by metal injection and a UV background as well
as by X-ray irradiation.
Our aim is to investigate whether star formation is significantly 
affected when the ISM of a proto-galaxxy enjoys different metallicities
and when a star forming cloud resides in the vicinity of a strong
X-ray source. We perform cosmological Enzo simulations with a detailed
treatment of non-zero metallicity chemistry and thermal balance. We also
perform FLASH simulations with embedded Lagrangian sink particles of a
collapsing molecular cloud near a massive, 10$^{7}$ M$_{\odot}$, black hole
that produces X-ray radiation.

We find that a multi-phase ISM forms for metallicites as small as $10^{-4}$
Solar at $z=6$, with higher ($10^{-2}Z_\odot$) metallicities supporting
a cold ($<100$ K) and dense ($>10^3$ cm$^{-3}$) phase at higher ($z=20$)
redshift.
A star formation recipe based on the presence of a cold dense phase leads
to a self-regulating mode in the presence of supernova and radiation feedback.
We also find that when there is strong X-ray feedback a collapsing cloud
fragments into larger clumps whereby fewer but more massive protostellar
cores are formed. This is a consequence of the higher Jeans mass in the
warm (50 K, due to ionization heating) molecular gas.
Accretion processes dominate the mass function and a near-flat, non-Salpeter IMF results.

\end{abstract}

\firstsection 
\section{Brief Context}

The presence of metals and X-rays in star forming regions affect processes like gas heating, ionization
degree and cooling time, and do so up to column densities of 10$^{24}$ cm$^{-2}$
that are typical of dense cores (e.g., Hocuk \& Spaans 2010; Meijerink \& Spaans 2005).
At the same time, star formation in the early universe and inside active galactic nuclei may depend
strongly on the ambient metallicity and radiation background as well as on mechanical feedback from
supernovae (e.g., Greif et al.\ 2010).

\begin{figure}[t]
\begin{center}
\begin{tabular}{lll}
\begin{minipage}{5.5cm}
 \includegraphics[width=2in]{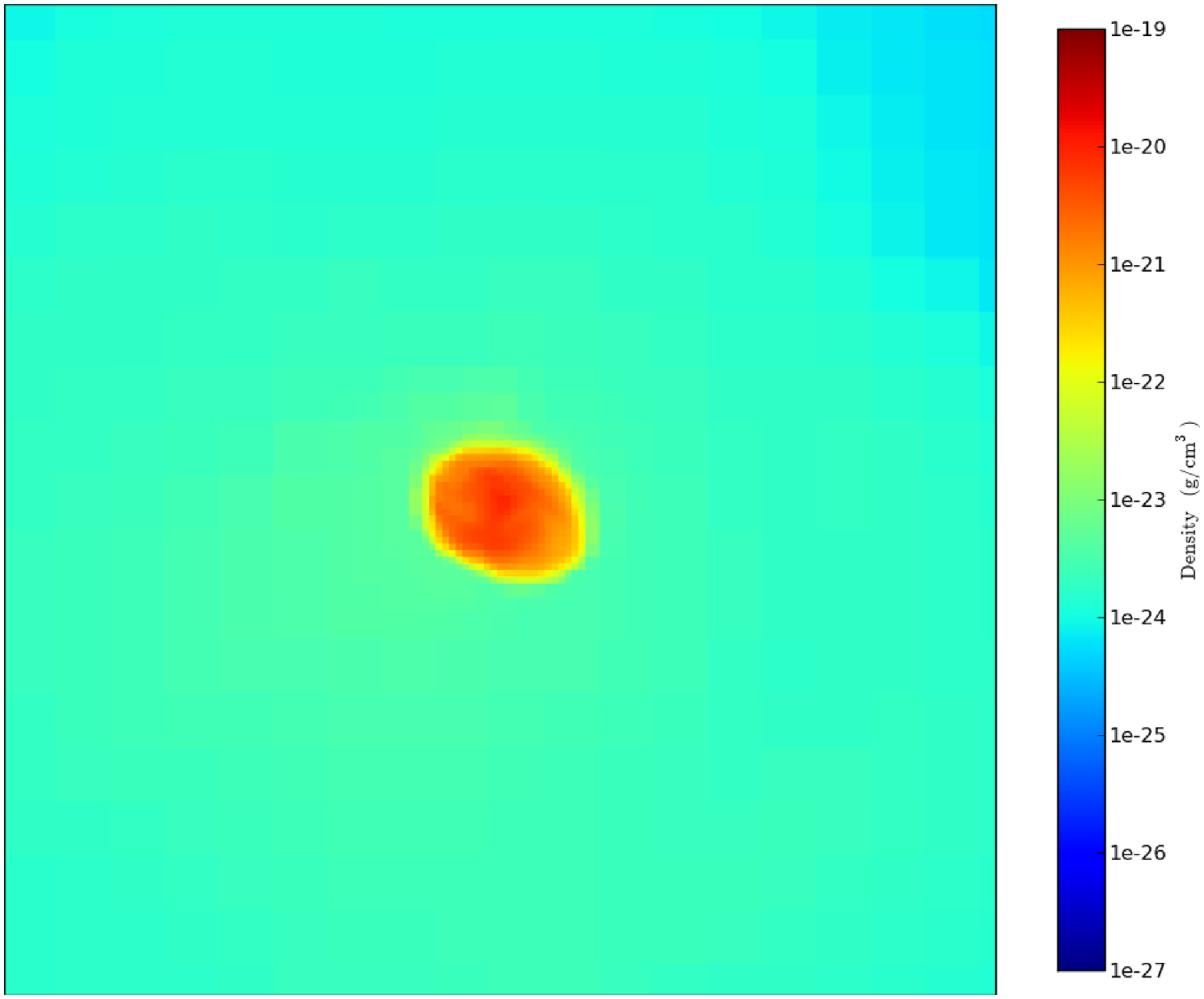}
\end{minipage}&
\begin{minipage}{5.5cm}
 \includegraphics[width=2in]{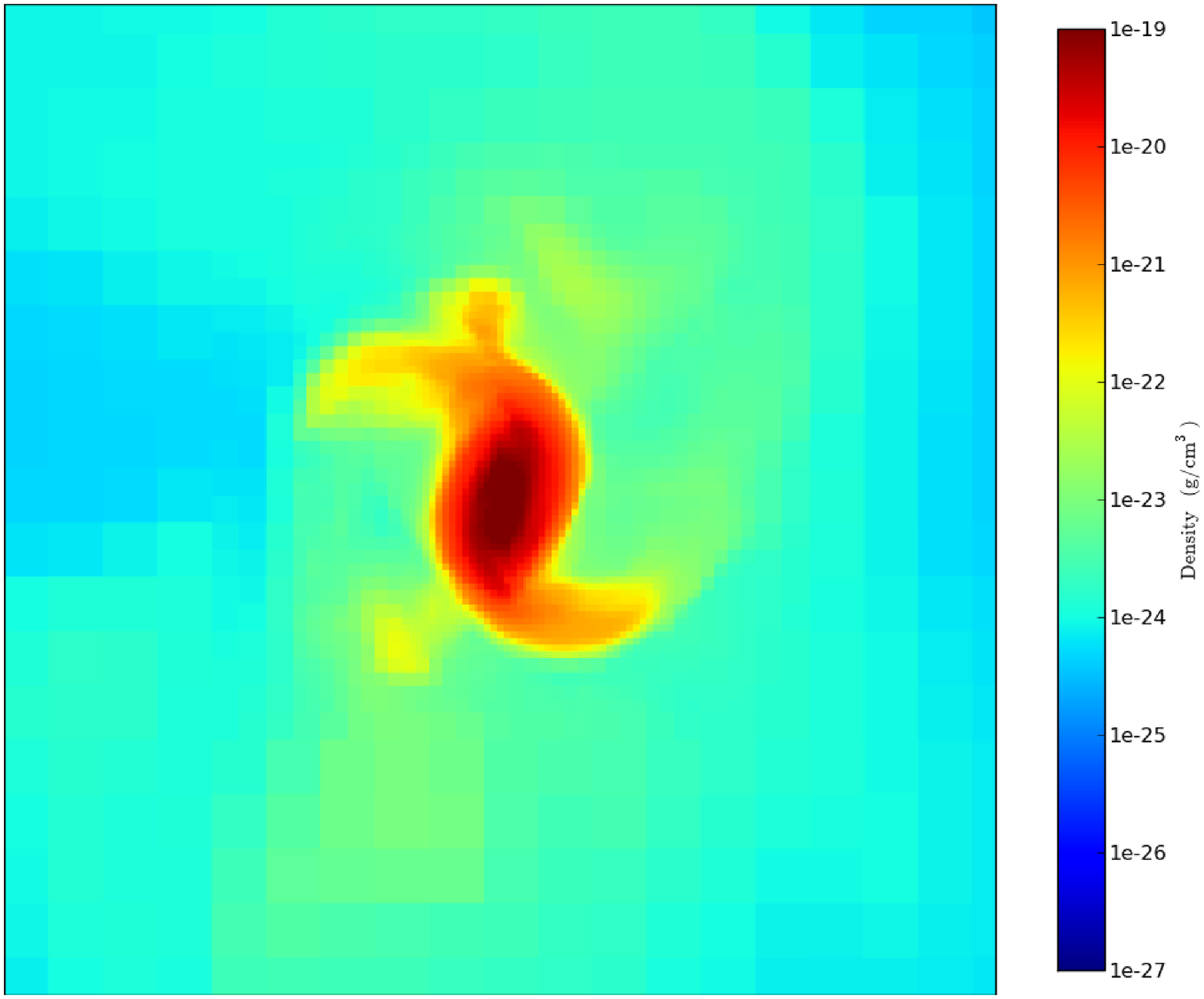}
\end{minipage}&
\begin{minipage}{5.5cm}
 \includegraphics[width=2in]{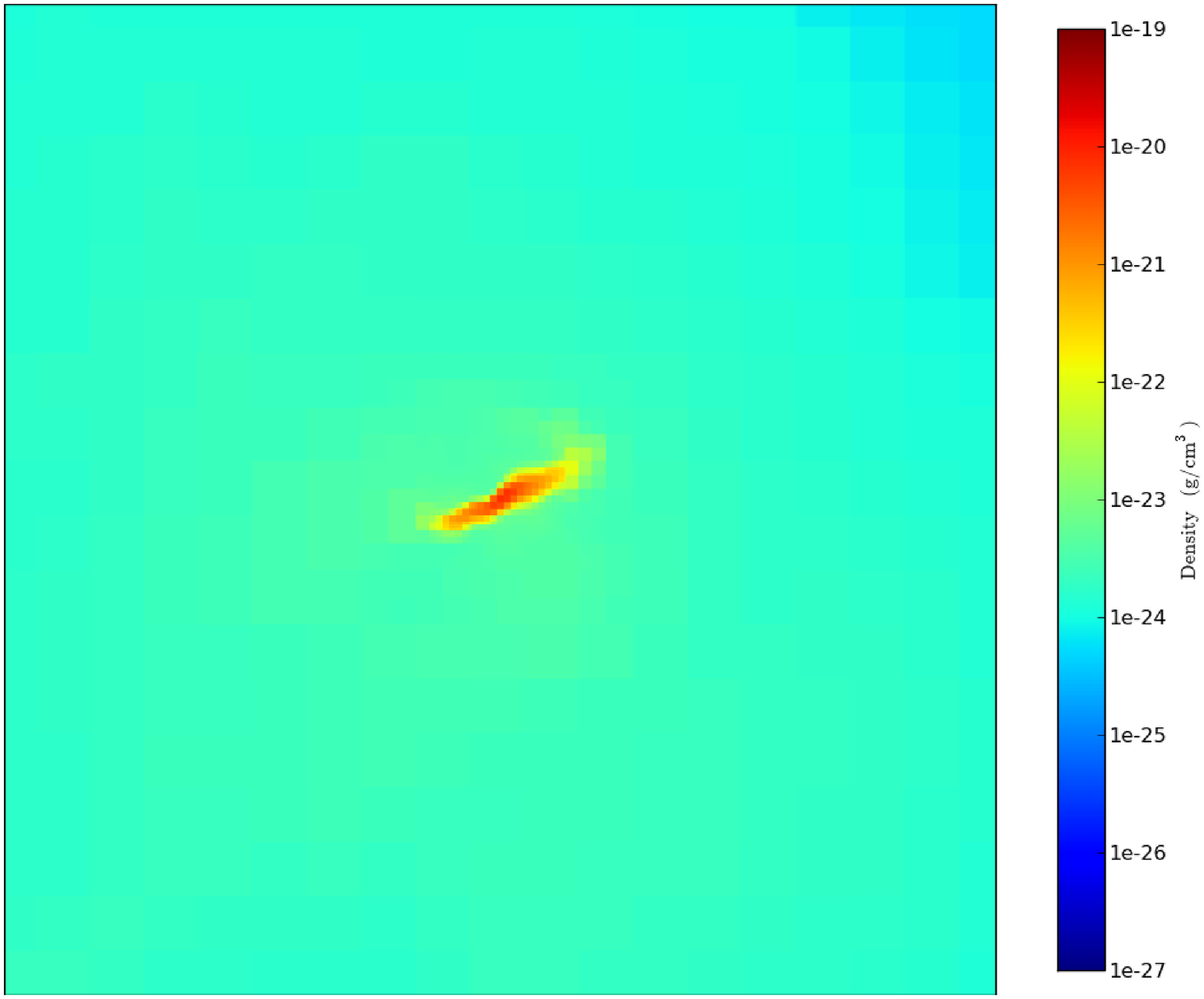}
\end{minipage} \\
\end{tabular}
 \caption{Density profiles of a minihalo at $z\approx 5.4$ for metallicities of
$Z/Z_\odot =10^{-4}$ (left), $Z/Z_\odot =10^{-2}$ (middle), and $Z/Z_\odot =10^{-2}$
with a $J_b=100J_{-21}$ radiation field (right).}
   \label{fig1}
\end{center}
\end{figure}

\section{Low-Metallicity Halos at High Redshift}

In the Enzo simulations we use $128^3$ grid cells on the top grid with three nested subgrids, each refining
by a factor of two. The box size of the simulations is 1-8 Mpc/h and a concordance cosmology is adopted.
We run simulations from redshift 99 to 5 for a pre-enriched ISM, using different cooling prescriptions
for metallicities of $10^-4$, $10^{-3}$, $10^{-2}$ and $10^{-1}$ Solar. In these simulations,
we have used a star formation recipe which derives from the metallicity dependent ISM phase
structure and we include UV radiation and supernova feedback from PopIII stars.
The metallicity dependent cooling is computed with the Meijerink \& Spaans (2005) UV chemistry
code and includes fine-structure emissions from carbon and oxygen as well as molecular lines from
species like H$_2$, CO and water. Heating is provided by adiabatic compression, dust grain photo-electric
emission and supernovae. All level populations are computed under statistical equilibrium.
The chemistry includes gas phase and grain surface formation of H$_2$ and HD (Cazaux \& Spaans
2004, 2009). CPU demands prevent global radiation transfer to be performed. Instead, we work
with a precomputed grid of cooling rates that augment the zero metallicity rates already
present in Enzo.

\begin{figure}[b]
\begin{center}
\begin{tabular}{cc}
\begin{minipage}{7.0cm}
 \includegraphics[scale=0.28]{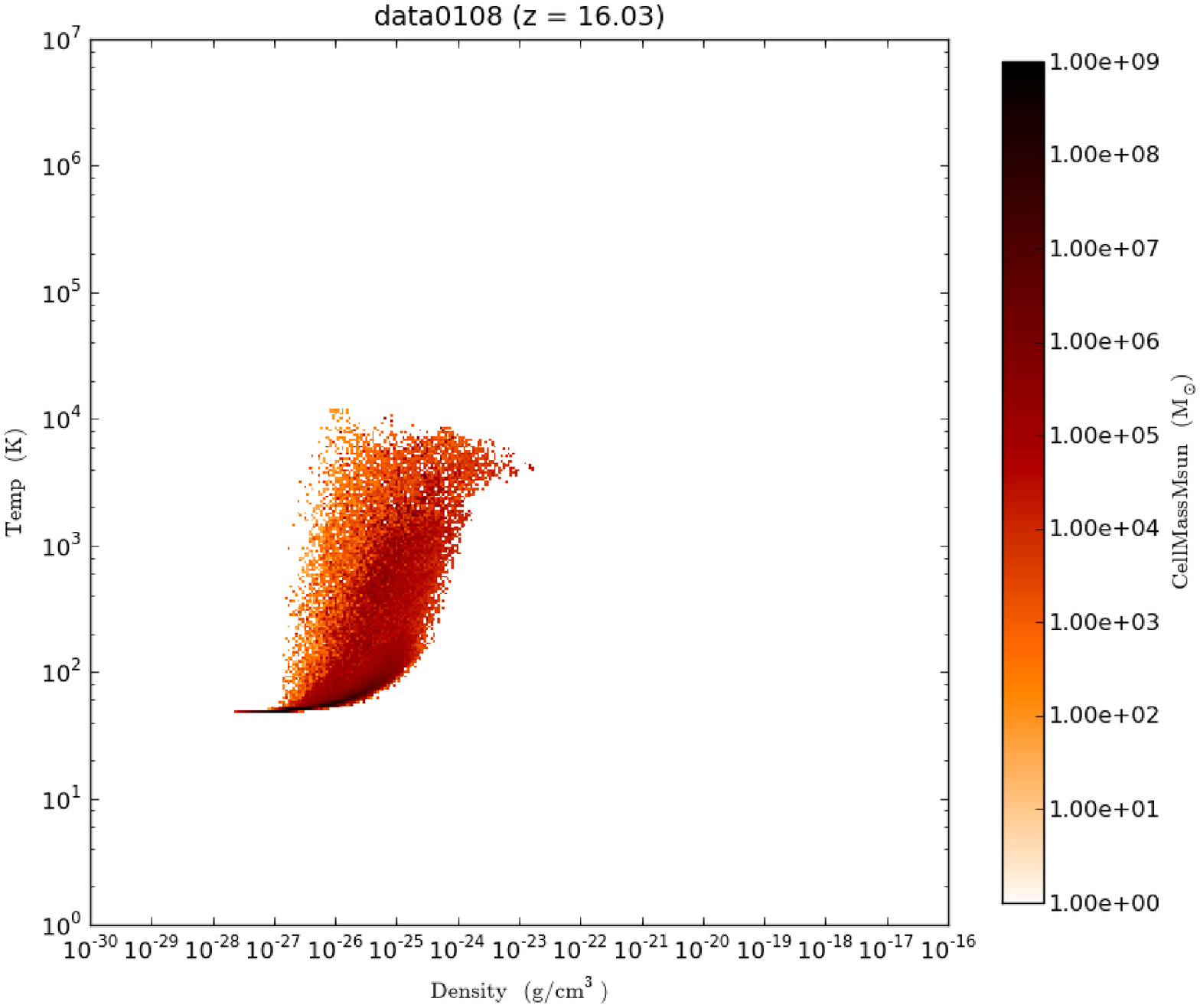}
\end{minipage}&
\begin{minipage}{7.0cm}
 \includegraphics[scale=0.28]{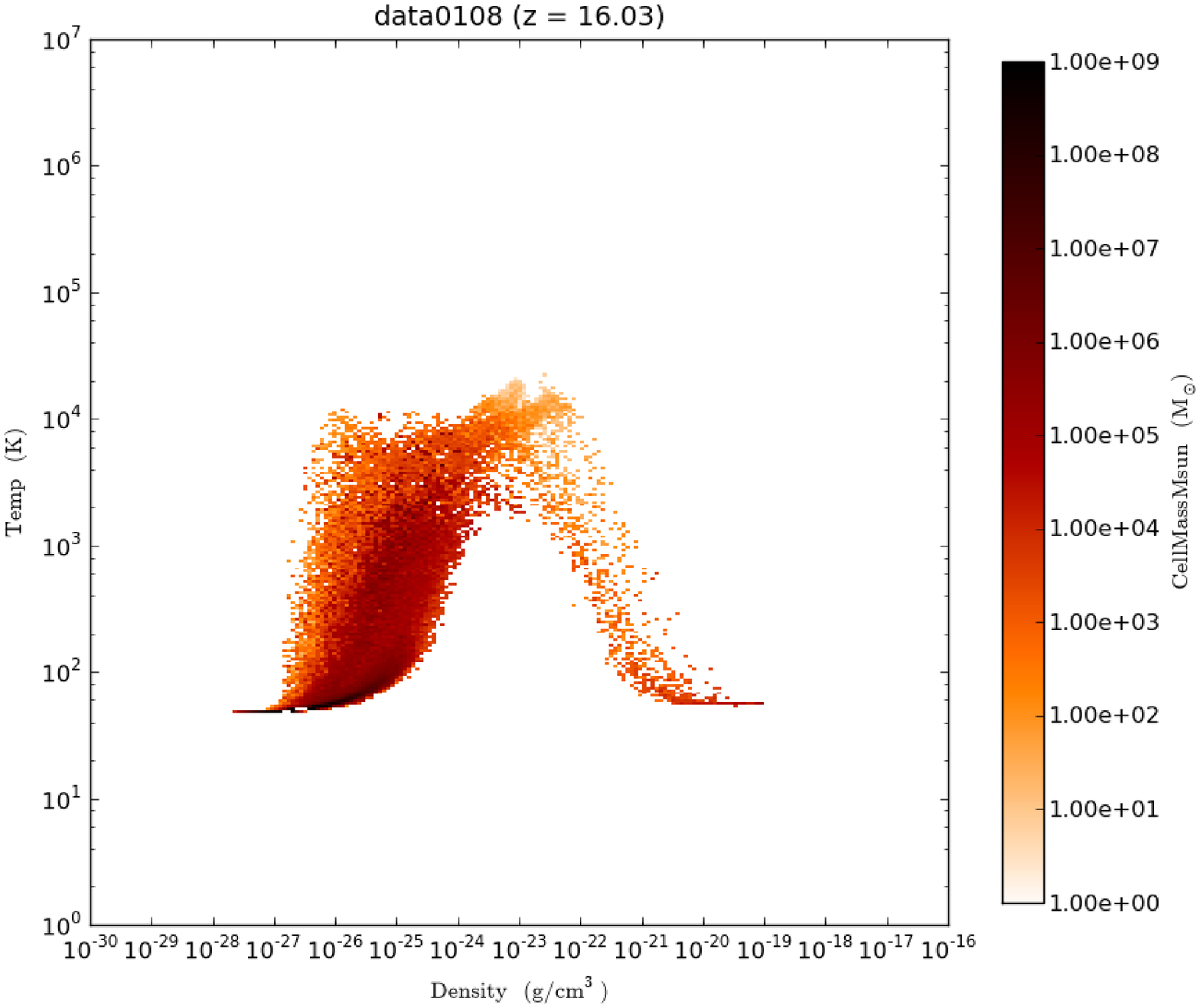}
\end{minipage} \\
\end{tabular}
 \caption{Density-temperature phase diagrams at $z\approx 16$ for $Z/Z_\odot =10^{-4}$ (left)
and $Z/Z_\odot =10^{-2}$ (right).}
   \label{fig2}
\end{center}
\end{figure}

A multi-phase ISM develops, at various times, once the metallicity
exceeds a value of about $10^{-4}$ Solar, consistent with Jappsen et al.\ (2009).
The key feature in this is the presence of a cold ($<100$ K) and dense ($>10^3$ cm$^{-3}$) phase.
The redshift at which a multi-phase ISM is established depends on the metallicity,
see figure 2 for $z\approx 16$ where. Also, the $Z/Z_\odot =10^{-2}$ case evolves faster
dynamically compared to the $Z/Z_\odot =10^{-4}$ case due to the shorter cooling time.
The higher metallicity halo is more compact and although mergers
are more violent it recovers faster than the $Z/Z_\odot =10^{-4}$ case (see figure 1, left
two panels).
In a run where we switch to $Z/Z_\odot =10^{-1}$ at redshift 12 we first form more stars due to
a stronger presence of a cold dense phase. Consequently, the minihalo experiences severe feedback effects.
Due to the efficient cooling ability of the metal-rich gas, the halo maintains its density structure but\
the cold dense phase is largely destroyed (see figure 3).

\begin{figure}[t]
\begin{center}
\begin{tabular}{ll}
\begin{minipage}{8.5cm}
 \includegraphics[scale=0.5]{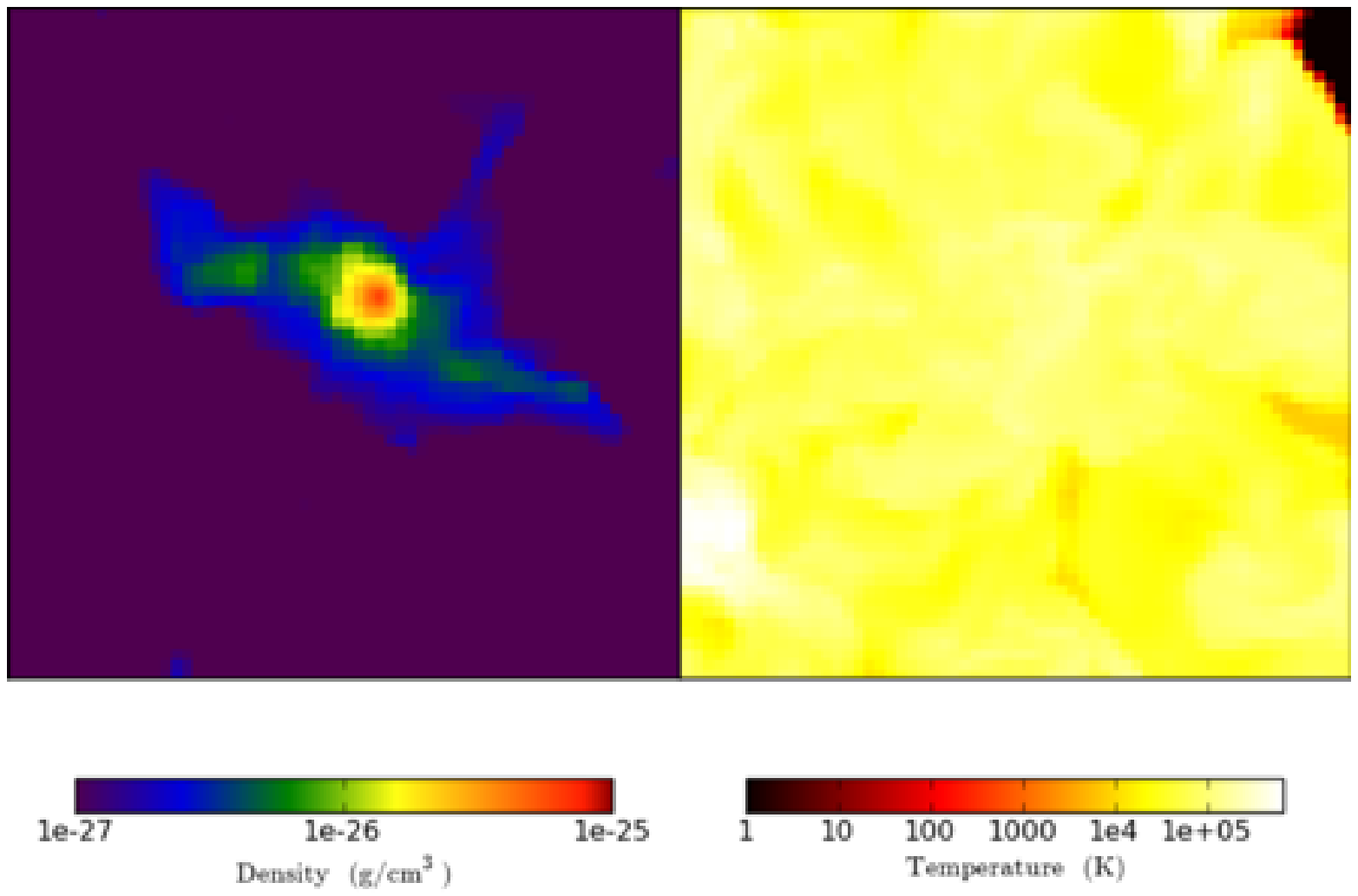}
\end{minipage}&
\begin{minipage}{8.5cm}
 \includegraphics[scale=0.5]{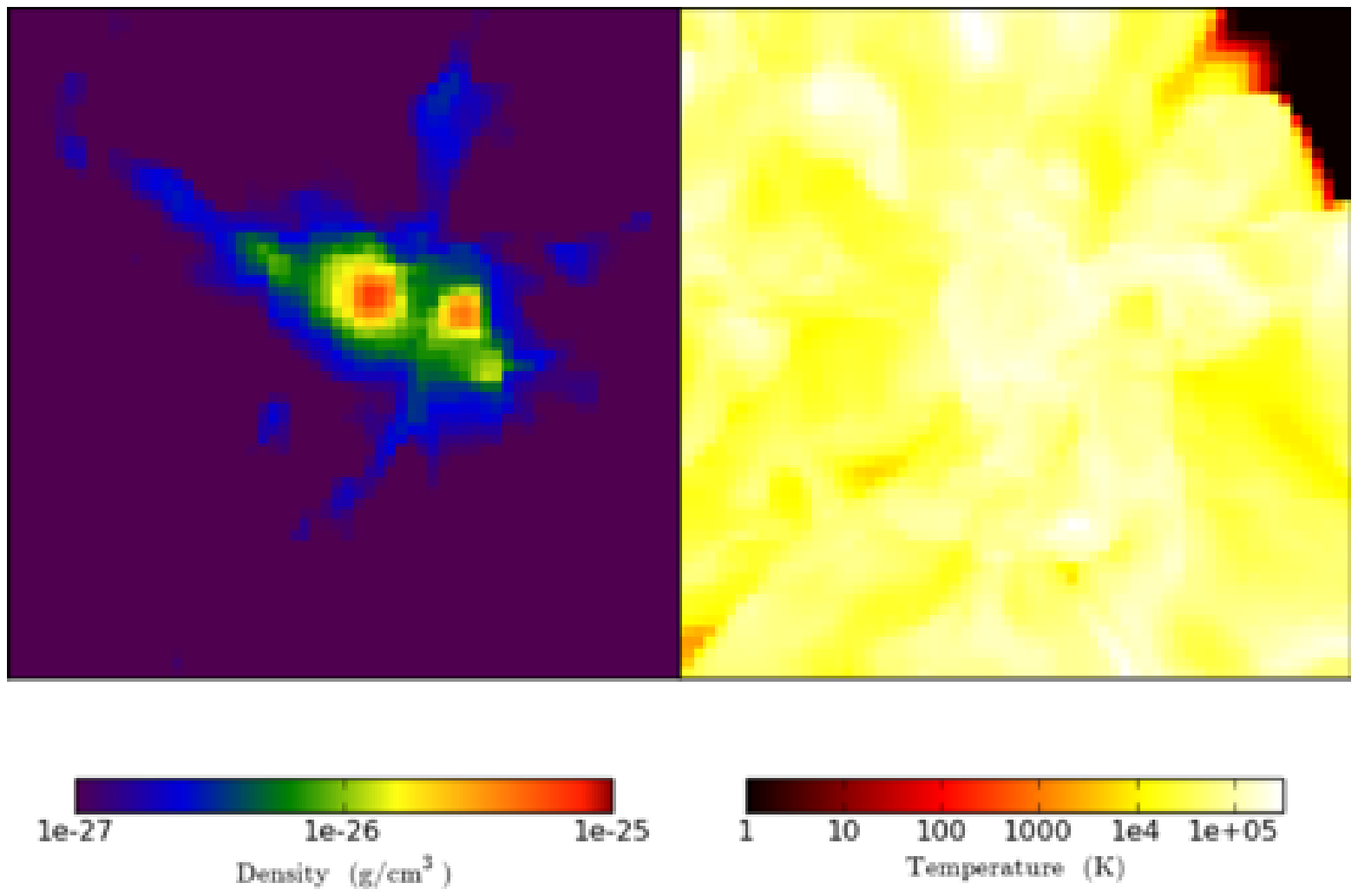}
\end{minipage} \\
\end{tabular}
 \caption{Density-temperature slices at $z\approx 6$ for star formation and feedback on runs for
$Z/Z_\odot =10^{-2}$ (left) and $Z/Z_\odot =10^{-2}$  increasing to $Z/Z_\odot =10^{-1}$ at $z=12$ (right).}
   \label{fig3}
\end{center}
\end{figure}

As a first attempt to include the effects of UV radiation feedback, we have performed runs
with a constant background radiation field of magnitude equal to the mean Galactic
value (about $J_b=100J_{-21}$). This background dissociates molecules like CO completely
and leaves little H$_2$ despite the self-shielding ability of the latter. It should be
emphasized that $J_b$ is merely chosen and that the transfer of radiation is strictly local.
Still, these first calculations do indicate that the cold dense phase is
fragile for metallicities of $Z/Z_\odot\le 10^{-2}$ (figure 1, right panel).
In all, supernova and radiation feedback act to diminish the presence of a cold
dense phase (at a given metallicity) and thus allow self-regulation for a star
formation recipe that is based on a cold and dense phase.

\section{X-ray Irradiation of a Cloud near a Black Hole}

In order to follow the effects caused by X-rays on the temperature of the gas, the X-ray
chemistry code of Meijerink \& Spaans (2005) is ported into FLASH. This code again includes
all relevant atomic and molecular cooling lines and carefully treats secondary ionizations
by fast electrons, Coulomb heating and ion-molecule/neutral-neutral reactions.
Given an X-ray flux, gas density, and global column density along the line of sight to the source, 
the code calculates the temperature and chemical abundances self-consistently. This output
is fed into the simulation at every iteration yielding a complete X-ray chemistry extension of FLASH.
The cloud is located at 10 pc from the black hole. A $k^{-4}$ turbulent spectrum is
imposed as well as Keplerian shear. The initial cloud mass is 800 M$_\odot$. The X-ray flux is
160 erg s$^{-1}$ cm$^{-2}$ with an $E^{-0.9}$ power-law distribution between 1 and 100 keV.
Sink particles are implemented following Federrath et al.\ (2010) and Krumholz et al.\ (2004).

\begin{figure}[t]
\begin{center}
 \includegraphics[width=3in]{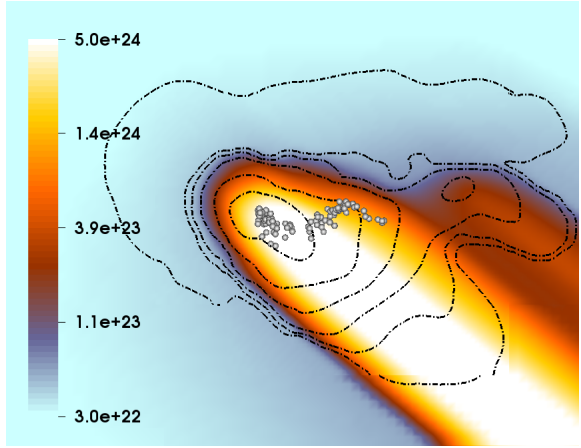}
 \caption{Column density (cm$^{-2}$) slice of a molecular cloud under the
impact of X-rays (at $t=t_{\rm ff}$). The colors represent the column density
along the line of sight to the black hole, which is located at the upper
left side. The density is shown in black contours. Contour levels are 1,
4, 16, 64, 256, 1024 $\times 10^4$ cm$^{-3}$. The white spheres
indicate the locations of the protostars.}
   \label{fig4}
\end{center}
\end{figure}

X-rays heat the cloud from one side, where the accreting black hole is located.
This increases the thermal pressure and causes the outer gas layers to expand and evaporate.
At the same time, the molecular cloud is compressed as an ionizing pressure flow travels
inwards. Finger-like shapes are formed, with a high density
head, and the gas that is lying in their shadows is well shielded, see figure 4.
The increased density and external thermal pressure caused the efficient
($\sim 30$\%) ionization heating then induces star formation.

\begin{figure}[t]
\begin{center}
\begin{tabular}{cc}
\begin{minipage}{8.5cm}
 \includegraphics[scale=0.5]{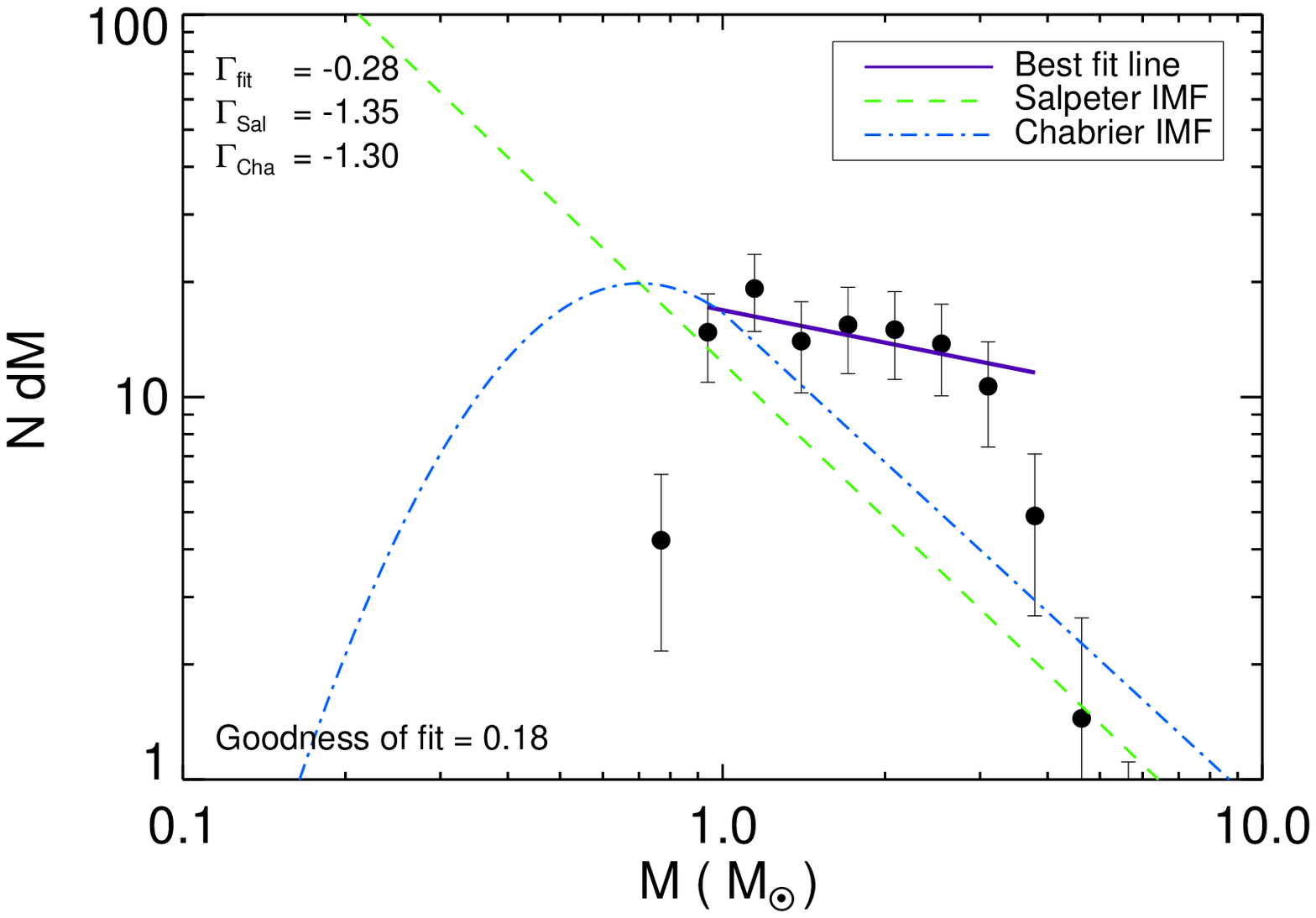}
\end{minipage}&
\begin{minipage}{8.5cm}
 \includegraphics[scale=0.5]{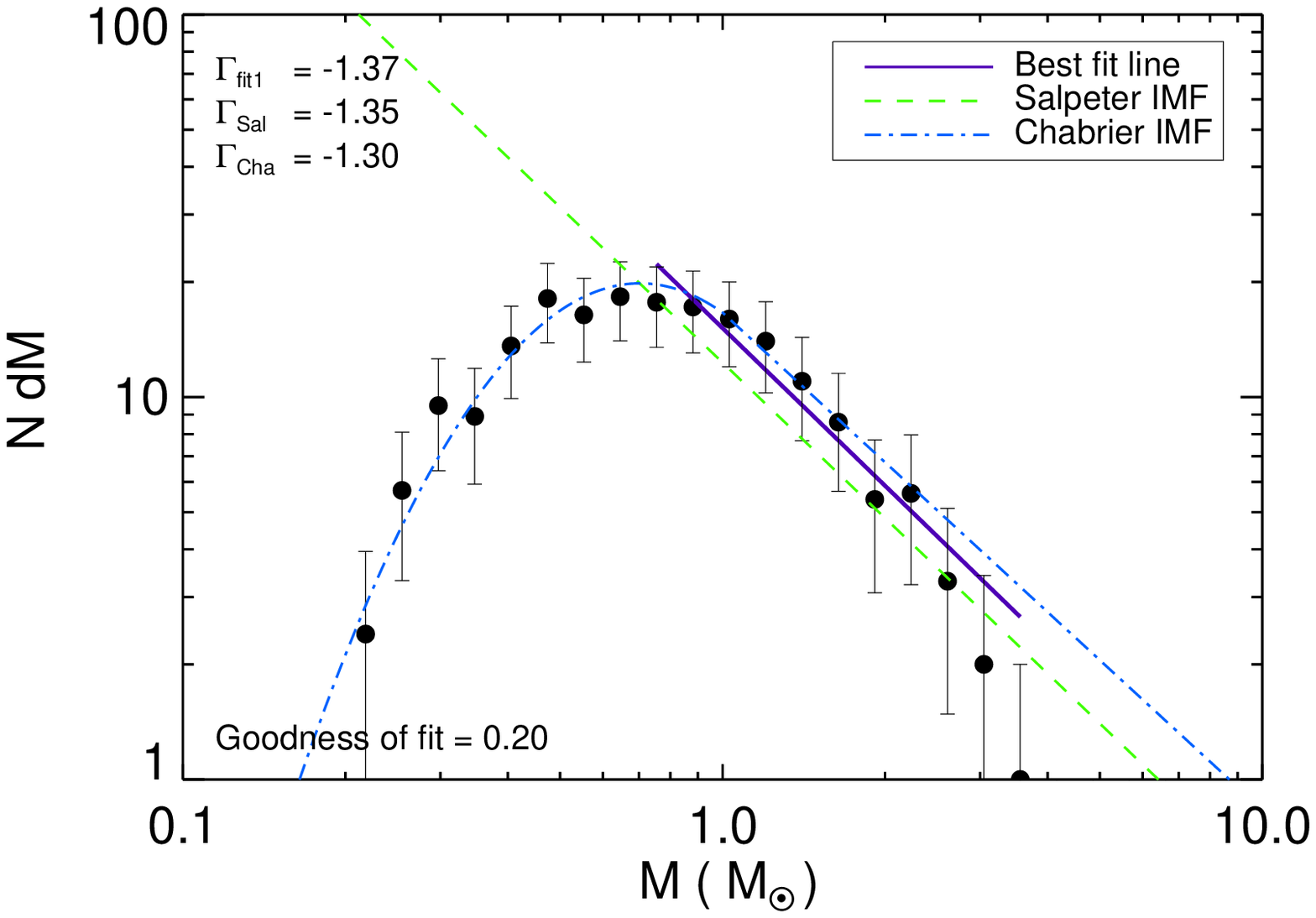}
\end{minipage} \\
\end{tabular}
 \caption{The IMF of two simulations at $t=t_{\rm ff}$.
Left  panel shows the IMF for the simulation with X-rays, and the right panel
shows the IMF for the simulation without X-rays.}
   \label{fig5}
\end{center}
\end{figure}

The simulation with X-rays results in a much flatter IMF slope than the one
without radiation, $\Gamma_{\rm X-ray}=-0.28$ after $10^5$ years 
of evolution (see figure 5). This is much flatter than the Salpeter slope
of -1.35 that we find for the simulation without X-rays, which has a
typical cloud temperature of 10 K due to cosmic ray heating only.
These trends are maintained at higher dynamical times,
$t=2t_{\rm ff}$. The main reason for the change in IMF comes from the
Jeans mass. The efficient X-ray heating yields temperatures in the
cloud of 50 K at densities as high $10^6$ cm$^{-3}$ and columns of
$10^{23-24}$ cm$^{-2}$. We find that the sink particles still grow in
mass after $2t_{\rm ff}$ and experience competitive accretion.

\end{document}